\documentclass[prd, preprint, nofootinbib, showpacs]{revtex4}

\usepackage{epsf,epsfig,subfigure,graphicx,amsmath,amssymb}
\usepackage{amsfonts}
\usepackage{latexsym}
\usepackage{color}

\newcommand{\dis}[1]{\begin{equation}\begin{split}#1\end{split}\end{equation}}
\newcommand{\be}{\begin{equation}}
\newcommand{\ee}{\end{equation}}
\def\bea{\begin{eqnarray}}
\def\eea{\end{eqnarray}}

\newcommand{\eq}[1]{Eq.~(\ref{#1})}

\newcommand{\VEV}[1]{\langle #1 \rangle}

\newcommand\gev{\,{\rm GeV}}

\newcommand\plin{p_{\rm lin}}
\newcommand\pcirc{p_{\rm circ}}

\begin{document}

\title{
 Quasar polarization with two axionlike particles
}

\author{Ki-Young Choi$^{1,2}$\footnote{kiyoung.choi@apctp.org},  Subhayan Mandal$^{1}$\footnote{smandal@apctp.org}, and Chang Sub Shin$^{1}$\footnote{csshin@apctp.org}}
\affiliation{${}^{1}$ Asia Pacific Center for Theoretical Physics, Pohang, Gyeongbuk 790-784, Republic of Korea }
\affiliation{${}^{2}$Department of Physics, POSTECH, Pohang, Gyeongbuk 790-784, Republic of Korea}

%

\begin{abstract}
Recently, it was shown, that  the absence of circular polarization of visible light from quasars, severely constrains the interpretation of axion like particles (ALPs) as a solution for the generation of the linear polarization. Furthermore, the new observation of the linear polarization in the radio wavelength from quasars, similar to the earlier observation performed in the optical band, makes the ALPs scenario inconsistent with at least one of the two observations. In this study we extend this scenario, to two axion like particles, one scalar and another pseudoscalar. We find, that the effects from scalar and pseudoscalar cancel out each other, thereby suppressing the circular polarization, while preserving consistent linear polarization, as observed in both visible and radio waves bands.

\end{abstract}


\preprint{APCTP-Pre2013 - 009} 

\vspace*{3cm}
\maketitle


\section{Introduction}
\label{intro}
The existence of the light scalars and pseudoscalars beyond the Standard Model is generic. One prominent example is, very light  axion, which was introduced to  solve the strong CP problem. The axion itself, being closely related to QCD, has a definite relation between its mass and decay constant. The interactions of axion to other particles are determined by specific models \cite{KSVZ79,DFSZ81}. 
Many other axionlike particles (ALPs) with very small mass are predicted, in the  supergravity \cite{maju} and superstring \cite{sen} theory. These ALPs can have a coupling to photons similar to the axion and may alter the astrophysical observations.
	
	In the last two to three decades \cite{maiani}, the axion-photon coupling and subsequent mixing or conversion scenarios has received a great deal of attention \cite{das}, both phenomenologically \cite{saha} and experimentally \cite{lamy}. This is of particular interest in astrophysics, where this mixing of photons with axion makes the universe transparent to the former \cite{csaki} and also changes their polarization properties \cite{hut}. It is also believed to be potentially responsible for effects like `Supernovae dimming' \cite{csaki} or `Large scale coherent orientation' \cite{hut} of the universe, through quasar polarization.

	It was found and confirmed that the quasar polarization vectors are not randomly oriented as naturally expected, but appear concentrated around one preferential direction on very large spatial scales~\cite{Hutsemekers:1998,Hutsemekers:2000fv,Jain:2003sg,Hutsemekers:2005iz}. 
	The polarization vectors appear coherently aligned over huge ($\sim$ Gpc) regions of the sky located
	at both low ($z\sim0.5$) and high ($z\sim1.5$) redshifts and characterized by different preferred directions of the quasar polarization. It seems that the mean polarization angle rotate with redshift at the rate of $\sim 30 {}^\circ$ per Gpc~\cite{Hutsemekers:2005iz}. In North Galactic hemisphere, it rotates clockwise and in South Galactic one it does counter-clockwise.

	It was suggested that the mixing between the light and ALPs in the background of magnetic field
can explain the alignment of polarization in the visible light from quasars~\cite{Jain:2002vx}.
However, the consequent observation of vanishing circular polarization~\cite{Hutsemekers:2010fw} is a problem for the hypothesis.
Therefore it is argued that the ALPs cannot explain the alignments and only constrain ALPs~\cite{Payez:2011sh}.
Furthermore recently it was reported that  the radio waves (8.4 GHz) show the similar coherent orientations of polarization from quasars~\cite{Tiwari:2012rr} by extending the first investigation with no alignment in Ref~\cite{Joshi:2007yf}. This further disfavors the need for ALPs as an explanations~\cite{Payez:2012rc} and gives new constraint on ALPs~\cite{Payez:2012vf}.

In this study, we shall explore the above mentioned polarization properties of distant quasars, not with the  single axion, but with the help of two generic ALPs: scalar and pseudoscalar. 
 In the existence of the external magnetic field, the scalar and pseudoscalar couple to the specific polarization direction of the light.
 The scalar couples to the orthogonal direction of the magnetic field while the pseudoscalar couples to the parallel direction only. Thus when the mixing of both scalar and pseudoscalar with light is very similar, the  circular polarization is cancelled by their mutual effect on photon. We show this using Stokes parameters applied to the light interacting with two generic ALPs, mediated by a background magnetic field. Therefor the existence of two ALPs are quite compatible with observations. We also find that  the alignment of the mixing angles suppresses the rotation of the orientation angle of linear polarization and is difficult to model the effect redshift dependent, as reported in the observations, over the cosmological length scales.

In section~\ref{sec:pol} we summarize the properties of the light from the quasars, in terms of the stokes parameters and describe the observational findings in terms of them, in three different categories. In the next section~\ref{sec:twoalps} we shall introduce two channel mixing of photon with ALPs one being of scalar type \& the other being of pseudoscalar type. Here in this section we modify the formulae for stokes parameters \cite{Payez:2012rc} generally given in the literature for single channel photon to ALPs mixing. We  also develop a new formula for linear polarization orientation angle dependence with distance (redshift), which has not been discussed before in single channel mixing literatures, on the very least, not for this type of models that employ - uniform magnetic field (\& medium). We , in section~\ref{sec:result} delineate our results to compare the observations with our model. We summarize in section~\ref{sec:discussion} \&  comment on the viability of ALPs vis-a-vis `Large scale coherent orientation' \cite{hut} observation on quasars. We demonstrate that previous objections raised disfavouring ALPs as a viable model for the (HUTSEMEKERS) effect may be nullified in this novel route.	
	
\section{The polarization of light from quasar}
\label{sec:pol}
Any light beam can be  characterized by $\vec{\mathcal E} (z,t) = {\mathcal E}_{\parallel} (z,t) \vec{e}_{\parallel} +  {\mathcal E}_{\perp} (z,t) \vec{e}_{\perp}$. Here $(\vec{e}_{\parallel} ,\vec{e}_{\perp} )$ define the plane transverse to the propagation direction of the light and  each are chosen to be parallel and perpendicular to the transverse magnetic field to the direction of the light. 
The polarization state of the light can be described by the Stokes parameters  defined as
\dis{
I(z) &= \VEV{\mathcal{I} (z,t) } = \VEV{\mathcal{E}_{\parallel} \mathcal{E}^*_{\parallel} +  \mathcal{E}_{\perp} \mathcal{E}^*_{\perp} },\\
Q(z) &= \VEV{\mathcal{Q} (z,t) } = \VEV{\mathcal{E}_{\parallel} \mathcal{E}^*_{\parallel} -  \mathcal{E}_{\perp} \mathcal{E}^*_{\perp} },\\
U(z) &= \VEV{\mathcal{U} (z,t) } = \VEV{\mathcal{E}_{\parallel} \mathcal{E}^*_{\perp} +  \mathcal{E}^*_{\parallel} \mathcal{E}_{\perp} } = 2Re \VEV{ \mathcal{E}_{\parallel} \mathcal{E}^*_{\perp}  },\\
V(z) &= \VEV{\mathcal{V} (z,t) } = \VEV{i(-\mathcal{E}_{\parallel} \mathcal{E}^*_{\perp} +  \mathcal{E}^*_{\parallel} \mathcal{E}_{\perp} )}= 2Im \VEV{ \mathcal{E}_{\parallel} \mathcal{E}^*_{\perp}  },
\label{stokes}
}
where the bracket denotes the averages over the exposure time.
Here $I$ denotes the intensity of the light and  $Q$ and $U$ describes the linear polarization and $V$ represents the circular polarization. The degree of the linear polarization and the degree of circular polarization are respectively 
\dis{
\plin = \frac{\sqrt{Q^2+U^2}}{I} = \sqrt{q^2+u^2},\qquad {\rm and}\qquad \pcirc = \frac{|V|}{I}=v , \label{plin}
}
and they contribute to the total degree of polarization,
\dis{
\qquad p_{\rm tot} = \frac{\sqrt{Q^2+U^2+V^2}}{I} = \sqrt{q^2+u^2 +v^2}.
}
Here  we used $q\equiv\frac{Q}{I}$, $u\equiv\frac{U}{I}$, and $v\equiv\frac{V}{I}$. 

It has been shown that the distribution of polarization position angles of the light from quasars is not random
and show correlations with distance. Below we summarize the features of the polarization of quasar light.

{\bf (I) Degreee of the linear polarization}\\
The light from distant quasars shows coherent orientations in the polarization both in visible and radio ($8.4$ GHz) waves.  The degree of  linear polarization is of the order of $1\%$. To explain the observation we require~\cite{Hutsemekers:2010fw}
\dis{
0.005  \lesssim  |p_{\rm lin} | \lesssim 0.02, \label{cond_plin}
}
both for visible and radio waves.

{\bf (II) Absence of the circular polarization}\\
Although the objects are highly linearly polarized, the circular polarization is not observed~\cite{Hutsemekers:2010fw}. This non-observation of circular polarization constrains $|p_{\rm circ}|$ by
\dis{
 \left| p_{\rm circ} \right| \lesssim 0.001.\label{cond_circ}
}

{\bf (III) Redshift dependence of the polarization direction}\\
Another interesting observation is the correlation of the polarization angles with the cosmological distances.
The observation shows the rotation of the angle is around~\cite{Hutsemekers:2005iz}
\dis{
30 ^{\circ} \quad {\rm per} \quad \rm{Gpc}.
\label{cond_angle}
}
However we note that  this shift in orientation with distance can be seen in the  restricted data set only (aligned to a particular direction), binned over large (z=0.5) distances, with number of entries in each bin being small~\cite{Hutsemekers:2005iz}. 

To date there is no known satisfactory explanation for these three observations. Recently there was a suggestion with axion like particles
mixing with light to explain the linear polarization. However the absence of circular polarization and the polarization in the radio waves disfavors the interpretation of ALP. But as we are working on this model, when there are two kinds of ALPs, scalar and pseudoscalar, the behavior is different.
In the next section, we will look for mechanisms with two ALPs to explain the above observations. 

\section{Two ALPs: scalar and pseudoscalar}
\label{sec:twoalps}

In this paper, we consider both scalar and pseudoscalar ALPs with different masses and couplings in general.
The Lagrangian density is given by
\begin{equation}
\mathcal{L} = -\frac{1}{4}F_{\mu\nu}F^{\mu\nu} + \frac{1}{2}\partial_\mu\phi\partial^\mu\phi - {1\over 2}m^2\phi^2 + \frac{1}{2}\partial_\mu\phi^{'}\partial^\mu\phi^{'} - {1\over 2}m^{'2}{\phi^{'}}^2  - \frac{1}{4}g \phi F_{\mu\nu}\tilde{F}^{\mu\nu}  -  \frac{1}{4}g'\phi^{'} F_{\mu\nu}F^{\mu\nu},
\label{Lagrangian} 
\end{equation}
where $m,m'$ and $g,g'$ are the masses and couplings of pseudoscalar and scalar respectively and $\tilde{F}^{\mu\nu}\equiv \frac12\epsilon^{\mu\mu\rho\sigma} F_{\rho\sigma}$ is the dual of the electromagnetic tensor.
The current limit on the coupling  is $g,g' < 6\times 10^{-11}\gev^{-1}$~\cite{Eidelman:1998jj}.


The equations of motion for the   scalar and pseudoscalar  are
\dis{
\partial^2\phi^{'} + m^{'2}\phi^{'}& = - \frac{1}{4}g_{\phi^{'}} F_{\mu\nu}F^{\mu\nu},\\
\partial^2\phi + m^2\phi &= - {1 \over 4} g_\phi\epsilon_{\mu\nu\rho\sigma}F_{\mu\nu}F^{\rho\sigma}.
}
Similarly, the equation of motion for the photon can be written as,
\begin{equation}
 	\Box\vec{E}+ \omega_p^2\vec{E}= g \vec{B}_{ext}\frac{\partial^2\phi}{\partial t^2} + g'\left(\vec{B}_{ext}\times\hat{n}\right)\frac{\partial^2\phi^{'}}{\partial t^2},
		\label{dalembert4}
\end{equation}
where $\vec{n}$ is the unit normal vector in the direction of $\vec{\nabla}\phi$ and we used $\vec{E}/c \ll \vec{B}_{ext}$ and $\vec{B} \ll \vec{B}_{ext}$. Here we added the term for the interaction with plasma with frequency $\omega_p$.

 
 In the existence of the external magnetic field, the scalar and pseudoscalar couple to the specific polarization direction of the light.
 The scalar couples to the orthogonal direction of the magnetic field while the pseudoscalar couples to the parallel direction. Therefore the mixing of ALPs with light are separated depending on the polarization of light.
 This makes the mass matrix a block diagonal one.
 The mixing angles between scalar and orthogonal polarization and between pseudoscalar and parallel polarization respectively are given by
\dis{
\theta^{'} = {1\over2}\arctan\left[\frac{2g'Bk}{-\omega_p^2+m^{'2}}\right],& \qquad \textrm{  for scalar}\\
\theta = {1\over2}\arctan\left[\frac{2gB\omega}{-\omega_p^2+m^2}\right], &\qquad \textrm{  for pseudoscalar},
\label{eq:mx-d}
}
where $B$ is the field strength of the  external magnetic field transverse to the direction of the light,  $\omega_p$ is plasma frequency, and $\omega$ and $k$ are the  frequency and wave number of the propagating light.
The eigenvalues for the masses are given respectively by
\begin{eqnarray}
\mu^{'2}_\pm &=& {1\over 2}\left[\left(\omega_p^2+m^{'2}\right) \pm \sqrt{\left(\omega_p^2-m^{'2}\right)^2 + \left(2g'Bk\right)^2}\right], \\
\mu^2_\pm &=& {1\over 2}\left[\left(\omega_p^2+m^2\right) \pm \sqrt{\left(\omega_p^2-m^2\right)^2 + \left(2gB\omega\right)^2}\right].
\label{masseigenstates}
\end{eqnarray}
For the initially vanishing ALP fields 
 \dis{
 \phi(0)=\phi'(0)=0, 
 }
we find  the field solutions as
\begin{eqnarray}
	\label{last-f}
	\phi^{'}(z) &=& {1\over 2}\sin(2\theta^{'})\left[~e^{iz\sqrt{\omega^2 - \mu^{'2}_+}} - ~e^{iz\sqrt{\omega^2 -\mu^{'2}_-}}\right] A_\perp(0), \nonumber \\
	A_\perp(z) &=& \left[\cos^2(\theta^{'}) e^{iz\sqrt{\omega^2 - \mu^{'2}_+}} + \sin^2(\theta^{'}) e^{iz\sqrt{\omega^2 -\mu^{'2}_-}}\right]A_\perp(0), \\
\phi(z) &=& {1\over 2}\sin(2\theta)\left[~e^{iz\sqrt{\omega^2 - \mu^{2}_+}} - ~e^{iz\sqrt{\omega^2 - \mu^{2}_-}}\right] A_\parallel(0), \nonumber \\
	A_\parallel(z) &=& \left[\cos^2(\theta) e^{iz\sqrt{\omega^2 - \mu^2_+}} + \sin^2(\theta) e^{iz\sqrt{\omega^2 - \mu^2_-}}\right]A_\parallel(0).
	\label{fields}
	\end{eqnarray}
Here $A_\perp$ and $A_\parallel$ are the orthogonal and parallel component of the electromagnetic potential
to the transverse external magnetic field.

Using \eq{last-f} and \eq{fields} we can obtain the evolutions of the Stokes parameters  using \eq{stokes} at a distance $z$ travelled  inside a magnetic field region for a plane wave described initially by $I_0,Q_0,U_0$, and $V_0$.
Those can be expressed as
\dis{
I(z) &= I_0 -\frac12(I_0+Q_0) \sin^22\theta \sin^2\left( \frac{\mu^2_+ -  \mu^2_- }{4\omega} z \right  )
-\frac12(I_0-Q_0) \sin^22\theta' \sin^2\left( \frac{\mu^{'2}_+ - \mu^{'2}_-}{4\omega}  z\right),\\
Q(z)&= I(I_0\longleftrightarrow Q_0),\\
U(z) &= U_0 \left\{ \cos^2\theta \cos^2\theta' \cos \left(\frac{\mu^2_+ - \mu^{'2}_+}{2\omega} z \right)  
+\sin^2\theta \sin^2\theta' \cos \left(\frac{\mu^2_- - \mu^{'2}_-}{2\omega}z \right)  \right.\\
&\left.\qquad +\cos^2\theta \sin^2\theta' \cos \left(\frac{\mu^2_+ - \mu^{'2}_-}{2\omega} z\right)  
+\sin^2\theta \cos^2\theta' \cos \left(\frac{\mu^2_- - \mu^{'2}_+}{2\omega} z\right)   \right\}\\
&-V_0\left\{ \cos^2\theta \cos^2\theta' \sin \left(\frac{\mu^2_+ - \mu^{'2}_+}{2\omega}z \right)  
+\sin^2\theta \sin^2\theta' \sin \left(\frac{\mu^2_- - \mu^{'2}_-}{2\omega} z\right)  \right.\\
&\left.\qquad +\cos^2\theta \sin^2\theta' \sin \left(\frac{\mu^2_+ - \mu^{'2}_-}{2\omega} z\right)  
+\sin^2\theta \cos^2\theta' \sin \left(\frac{\mu^2_- - \mu^{'2}_+}{2\omega} z\right)   \right\}, \\
V(z) &= U(U_0 \rightarrow V_0, V_0\rightarrow - U_0),
\label{stokesalps}
}
where we used the condition $\omega^2 \gg \mu^2_\pm$, $\mu^{'2}_\pm$.
As one can see, the mixing angles $\theta$ and $\theta'$ controls the magnitude of the polarization and the 
mass differences are responsible for the details of the propagation.

From the above Stokes parameters, one can readily see that  mixings of ALPs with photons inside the external magnetic fields changes the polarization of light and generates the phenomena called  'dichroism' and 'birefringence'. 

For initially unpolarized light ($Q_0=U_0=V_0=0$),
 the mixing generates non-vanishing polarization in $Q(z)$  due to the selective absorption of the polarization (dichroism) by ALPs while the circular polarization still remains vanishing, $U(z)=V(z)=0$. 
Even with partially polarized light ($Q_0\neq0$), the vanishing $U_0$ and $V_0$ do not make any circular polarization. 
However in general there is small non-vanishing $U_0$ and the mixing induces the interchange between the linear and circular polarization (birefringence) resulting in non-zero $V(z)$.   

\section{Result}
\label{sec:result}
In general, a partially polarized beam can be decomposed by the sum of fully polarized beam and of unpolarized one.  In this case the  the Stokes parameters are additive with these two beams.
Therefore in this section, We consider the light from quasar in two cases, one with initially unpolarized light and the other linearly polarized.

\subsection{Initially unpolarized light}
First we consider the degree of linear polarization in the case of initially unpolarized light beams. Thus we use 
\dis{
I_0\neq0, \qquad Q_0=U_0=V_0=0.
}
For initially unpolarized light, the circular polarization is not induced, however, the linear polarization is generated and the degree of the linear polarization  in \eq{plin} becomes
\dis{
p_{\rm lin} (z) = \frac{\left| \frac12 \sin^22\theta \sin^2\left( \frac{\Delta\mu^2 z }{4\omega}\right) -\frac12 \sin^22\theta' \sin^2\left( \frac{\Delta\mu^{'2} z }{4\omega} \right)\right|}{1-\frac12 \sin^22\theta \sin^2\left( \frac{\Delta\mu^2 z }{4\omega}\right) -\frac12 \sin^22\theta' \sin^2\left( \frac{\Delta\mu^{'2} z }{4\omega} \right)},
}
where 
\dis{
\Delta\mu^2\equiv \mu_+^2- \mu_-^2,\qquad \rm{ and}\qquad  \Delta\mu^{'2}\equiv \mu_+^{'2}- \mu_-^{'2}.
}
The mixing cannot produce the polarization given in~\eq{cond_plin} both in the visible and radio waves even with two ALPs, as similar to the case with single ALP~\cite{Payez:2012rc}.  Since the maximum polarization depends on the mixing angle and thus  on the frequency, the linear polarization have big difference between visible and radio waves, where the ratio of frequency between visible and radio wave are of an  order of $10^{5}$.  Therefore the degrees of polarization are different by an order of $10^{10}$. This implies that the observed polarization both in the visible and radio come  mainly from the intrinsic origin.

\subsection{Initially  polarized light}

We consider the initially polarized light beams with no circular polarization initially
\dis{
q(0)=0,\qquad u(0) =0.01, \qquad \rm{and}\qquad v(0) = 0.
\label{initialcond}
}
However the mixing with ALPs generate non-zero circular polarization due to the birefringence effect as in \eq{stokesalps}, which can easily violate the observation of the absence of circular polarization in \eq{cond_circ}.
However with two ALPs, in the following two cases,  we can find parameter region that the effects from scalar and pseudoscalar  cancel each other and result in the negligible circular polarization. 

{\bf 1) $m,m' \ll \omega_p$}\\
When the masses of the ALPs are much smaller than the plasma frequency, the mass eigenstates in \eq{masseigenstates} are 
\dis{
\mu^2_\pm& \simeq \frac12\omega_p^2\left(1\pm \frac{1}{\cos2\theta}\right),\\
\mu^{'2}_\pm& \simeq \frac12\omega_p^2\left(1\pm\frac{1}{\cos2\theta'}\right).
}
Therefore the absence of circular polarization in \eq{cond_circ}
is satisfied by the condition
\dis{
\sin (\theta+\theta')\, \Delta\theta  \lesssim \frac{10^{-3} 4\omega }{\omega_p^2 z u(0)} \simeq 10^{-3},
}
where $\Delta \theta \equiv \theta - \theta' \ll 1$. It is easy to find that  $q(z)$ is highly suppressed in this condition.
However $u(z)$ is always of the order of $u(0)$ and thus the degree of linear polarization is $\plin \simeq u(0)=0.01$.

{\bf 2) $m,m' \gtrsim \omega_p$}\\
In this case, to suppress the circular polarization in~\eq{stokesalps}, we need additional condition of the degenerate masses as well as the same order of mixing angles,
\dis{
\sin\theta \simeq \sin\theta' ,  \qquad {\rm and} \qquad m\simeq m'.
}
 

%
\begin{figure}[!t]
  \begin{center}
  \begin{tabular}{cc}
   \includegraphics[width=0.5\textwidth]{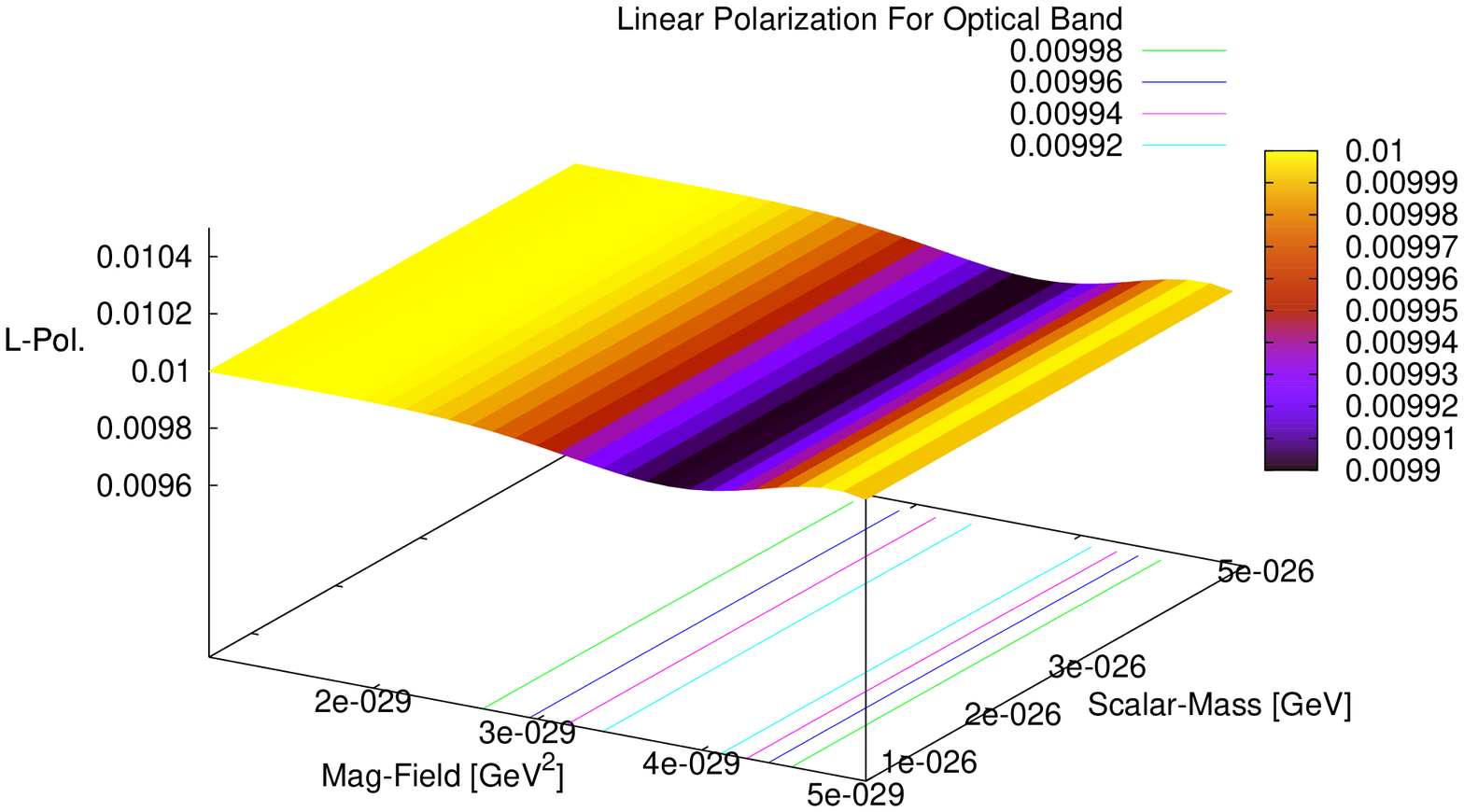}
  ~&~ 
     \includegraphics[width=0.5\textwidth]{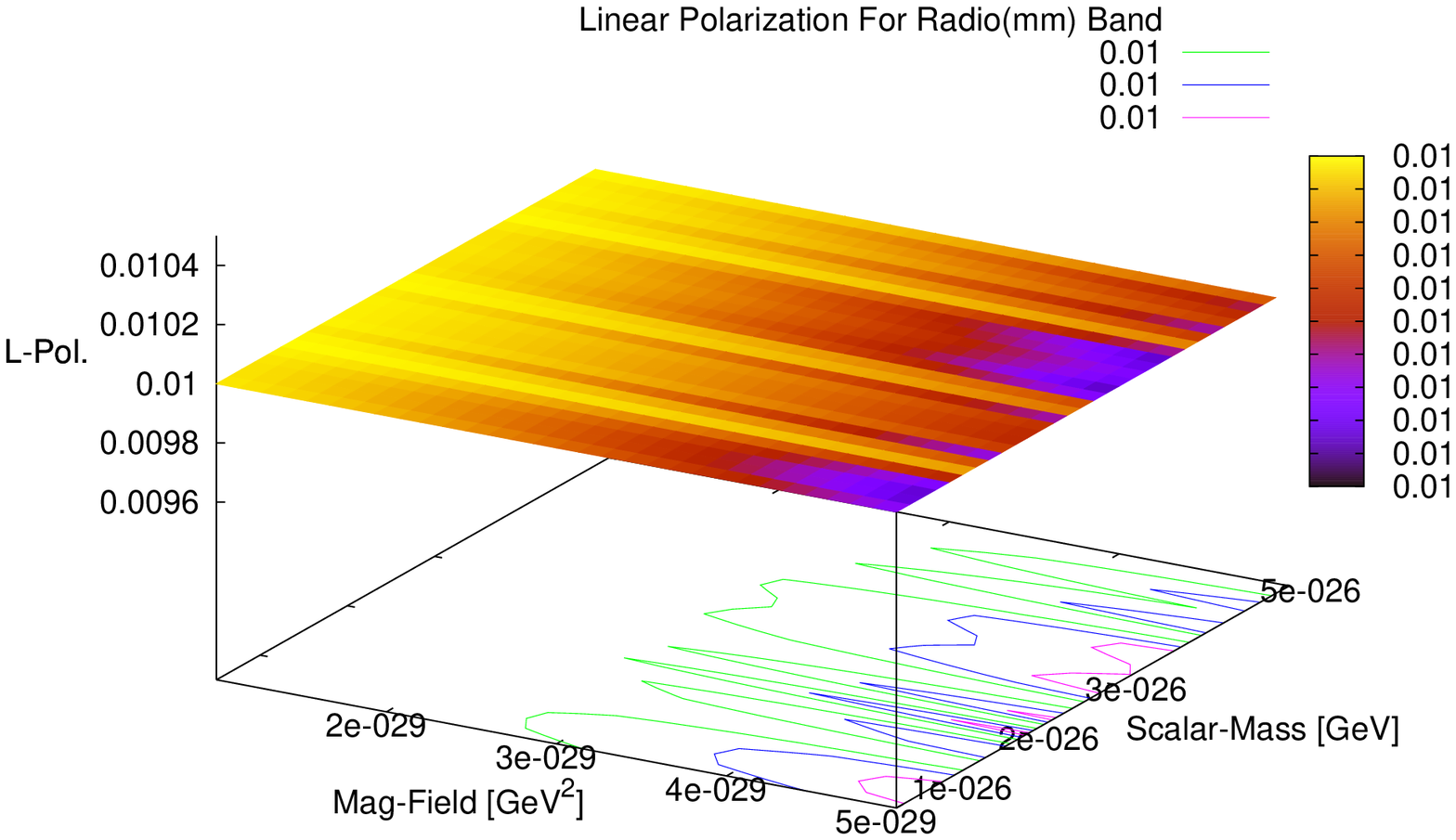}\\
   \includegraphics[width=0.5\textwidth]{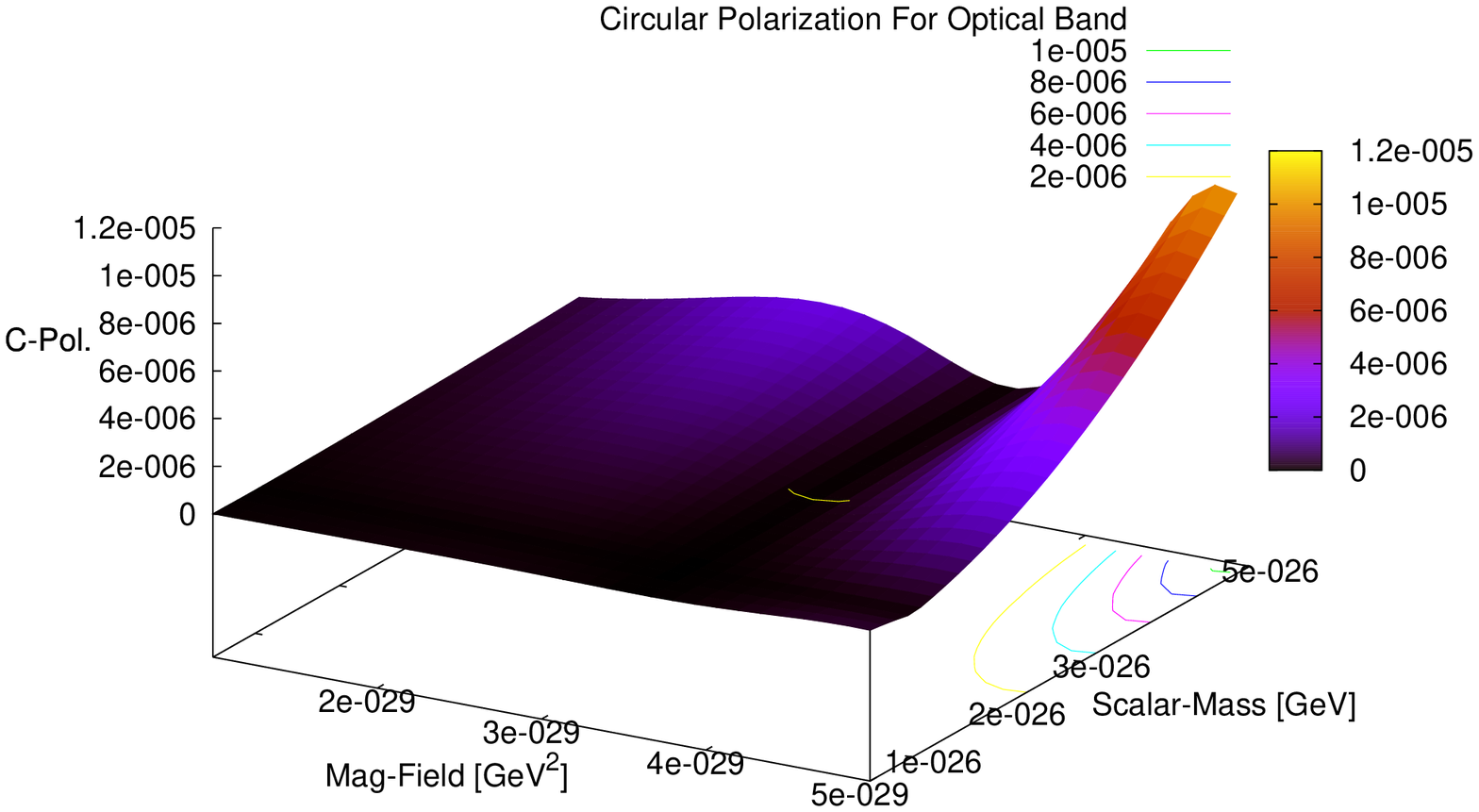}
  &
     \includegraphics[width=0.5\textwidth]{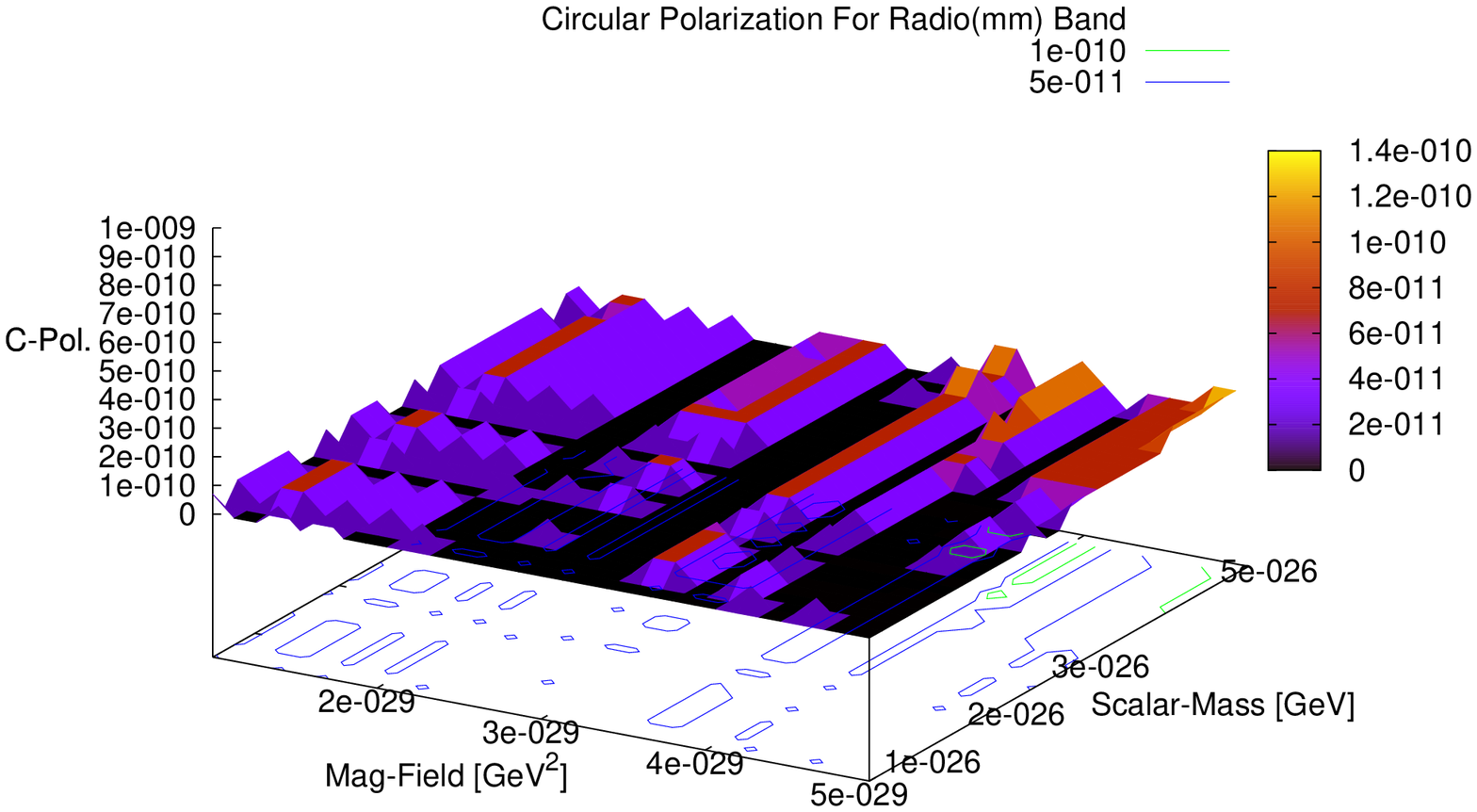}
  \end{tabular}
  \end{center}
  \caption{Case 1) : The contour plot of the degree of the linear (upper windows) and circular (lower windows) polarization for optical (left) and radio (right) waves respectively in the plane of the external magnetic field, $B$, and the scalar mass $m'$, which is varied for the fixed pseudoscalar mass  $m=10^{-26} \gev$.
  The plasma frequency has been chosen to be $\omega_p=4\times 10^{-24}\gev$.  The distance we used is  
  $1.28\,\rm{Gpc}=2\times 10^{41}\gev^{-1}$. As a initial values, we used $u(0)=0.01$ and $q(0)=v(0)=0$.  We consider that both the couplings are the same $g=g'=10^{-11}\gev^{-1}$. Here $m,m'\ll \omega_p$ and  $\sin\theta\simeq\sin\theta'\simeq -2.16\times 10^{-6}$ with $\Delta \theta \simeq 10^{-9}$.     
 }
  \label{fig:pol}
\end{figure}

In figure~\ref{fig:pol}, we show the contour plot of the linear  (upper) and circular (lower) polarization for the visible (left) and radio (right) waves  in the plane of the external magnetic field and the scalar mass for fixed pseudoscalar mass. 
The frequency of visible and radio waves are
$\omega_V = 2.5\times 10^{-9} \gev$  and  $\omega_R = 3.474\times 10^{-14} \gev$  which correspond to the wavelength $500\,\rm{nm} $ and $ 3.57\, \rm{cm}$ respectively.
We use the typical size of the   magnetic field strength of
$ 1 \rm{n} G = 1.95\times10^{-29}\, \mathrm{GeV}^2$ 
in the cosmological scale of $1\,\rm{Gpc}$~\cite{saha}.
Actually the exact value of the magnetic field strength is not an issue since the the coupling is always involved with magnetic field together.   
  The plasma frequency has been chosen to be $\omega_p=4\times 10^{-24}\gev$, though this is not a critical issue in our study. 
    We consider that both the couplings to be the same $g_\phi=g_{\phi'}= 10^{-11}\gev^{-1}$. 
    We fixed the pseudoscalar mass to be $m=10^{-26} \gev$ but  varied the other scalar mass $m'$ satisfying  $m,m'\ll \omega_p$.  For those parameters, we find that $\sin\theta\simeq\sin\theta'\simeq -2.16\times 10^{-6}$ with $\Delta \theta \simeq 10^{-9}$.     
    
In the upper windows of the figure~\ref{fig:pol}, we can see that the  degree of the linear polarization is
order of $0.01$ for both visible and radio waves, however the circular polarization is suppressed less than around $10^{-5}$ and $10^{-9} $ for visible and radio respectively. The high suppression in the radio frequency is because the mixing angles are much smaller due to the small frequency as in~\eq{eq:mx-d}.
Here we can see  that it is possible to nullify the circular polarization for wide range of parameters for the case 1) with $m,m'\ll \omega_p$ without resorting to the restrictive resonance condition.

\begin{figure}[!t]
  \begin{center}
  \begin{tabular}{cc}
   \includegraphics[width=0.5\textwidth]{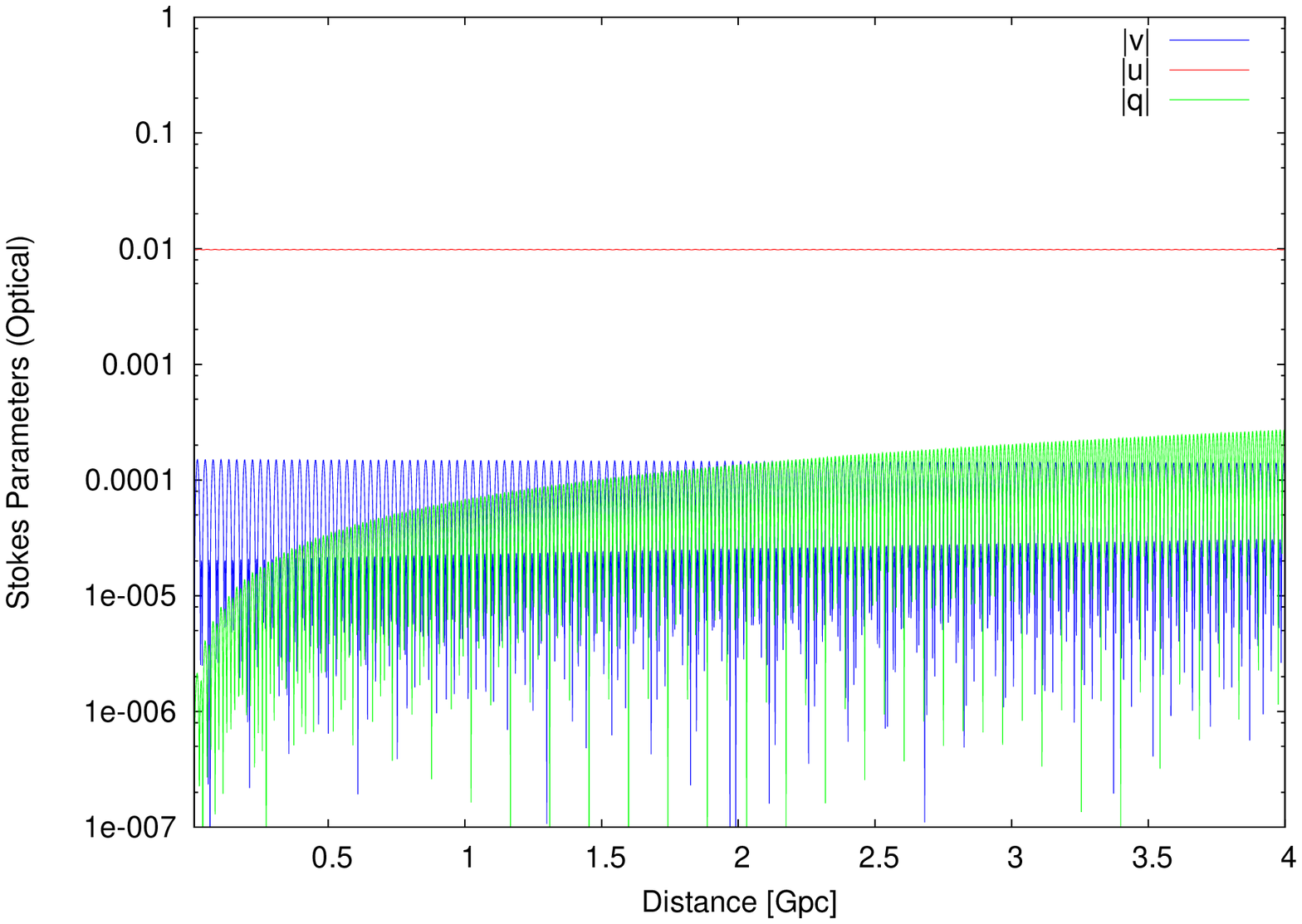}
  & 
     \includegraphics[width=0.5\textwidth]{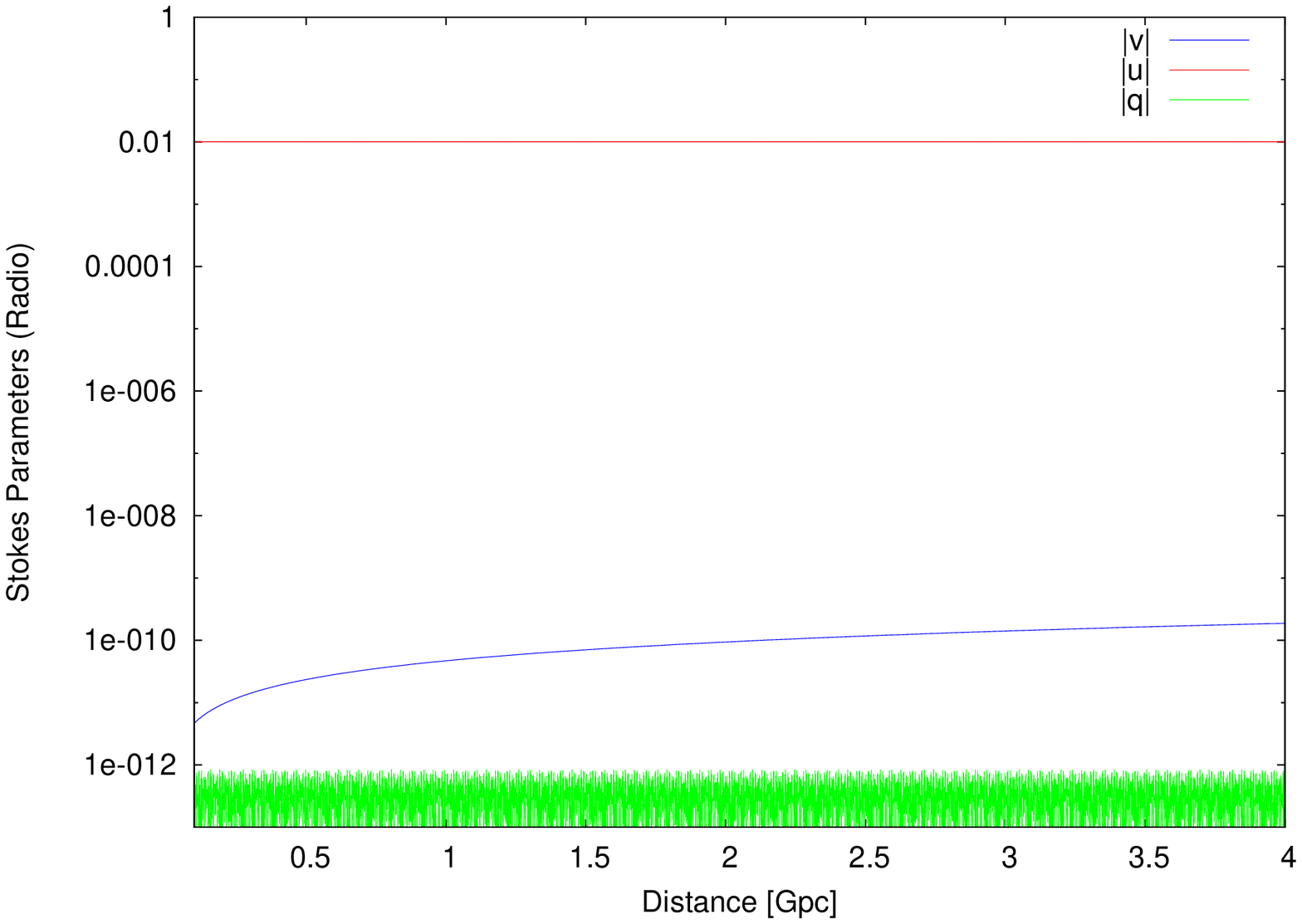}
  \end{tabular}
  \end{center}
  \caption{  The Stokes parameters $q,u$, and $v$ as a function of the distance for optical (left) and radio waves (right). We used the same parameters as used in figure~\ref{fig:pol} with $m'= 5\times 10^{-26} \gev$ and $B=2\times 10^{-29}\gev^2$.}
  \label{fig:poldistance}
\end{figure}
In figure~\ref{fig:poldistance} we show the evolution of the normalized Stokes parameters as a function of distance for the same parameters used in the figure~\ref{fig:pol} for the fixed values $m'= 5\times 10^{-26} \gev $ and $B=2\times 10^{-29}\gev^2$.

\begin{figure}[!t]
  \begin{center}
   \includegraphics[width=0.5\textwidth]{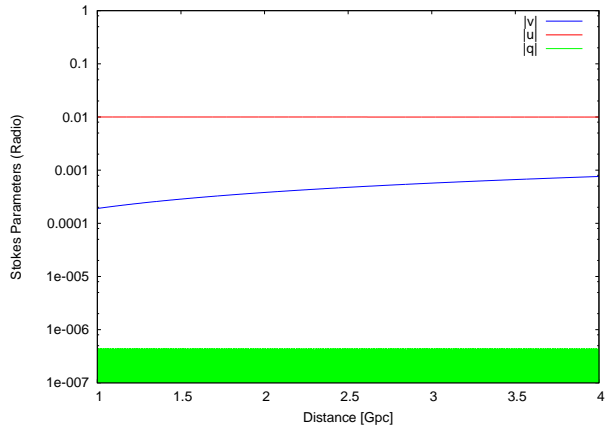}
  \end{center}
  \caption{ Case 2) :  The Stokes parameters $q,u$, and $v$ as a function of the distance for optical (left) and radio waves (right). Here We used the  parameters  $m= 4.0\times 10^{-23} \gev $,  $m'= 4.1\times 10^{-23} \gev$,  $g=g' = 3\times 10^{-11} \gev$ and $B=2\times 10^{-29}\gev^2$.}
  \label{fig:poldistance-large}
\end{figure} 

In figure \ref{fig:poldistance-large}, we show an example of case of 2) with  $m= 4.0\times 10^{-23} \gev $,  $m'= 4.1\times 10^{-23} \gev$, $g=g' = 3\times 10^{-11} \gev$ and $B=2\times 10^{-29}\gev^2$. That corresponds to the  mixing, $\sin\theta = \sin\theta'=  9\times 10^{-4}  $.  Here we only put the optical band stokes parameters, since, in the radio band the cirucalr polarization is very low and the $|q|$ polarization is indistinguishabe from zero. 
\subsection{Redshift dependence of the polarization angle}
\begin{figure}
  \begin{center}
  \begin{tabular}{cc}
   \includegraphics[width=0.5\textwidth]{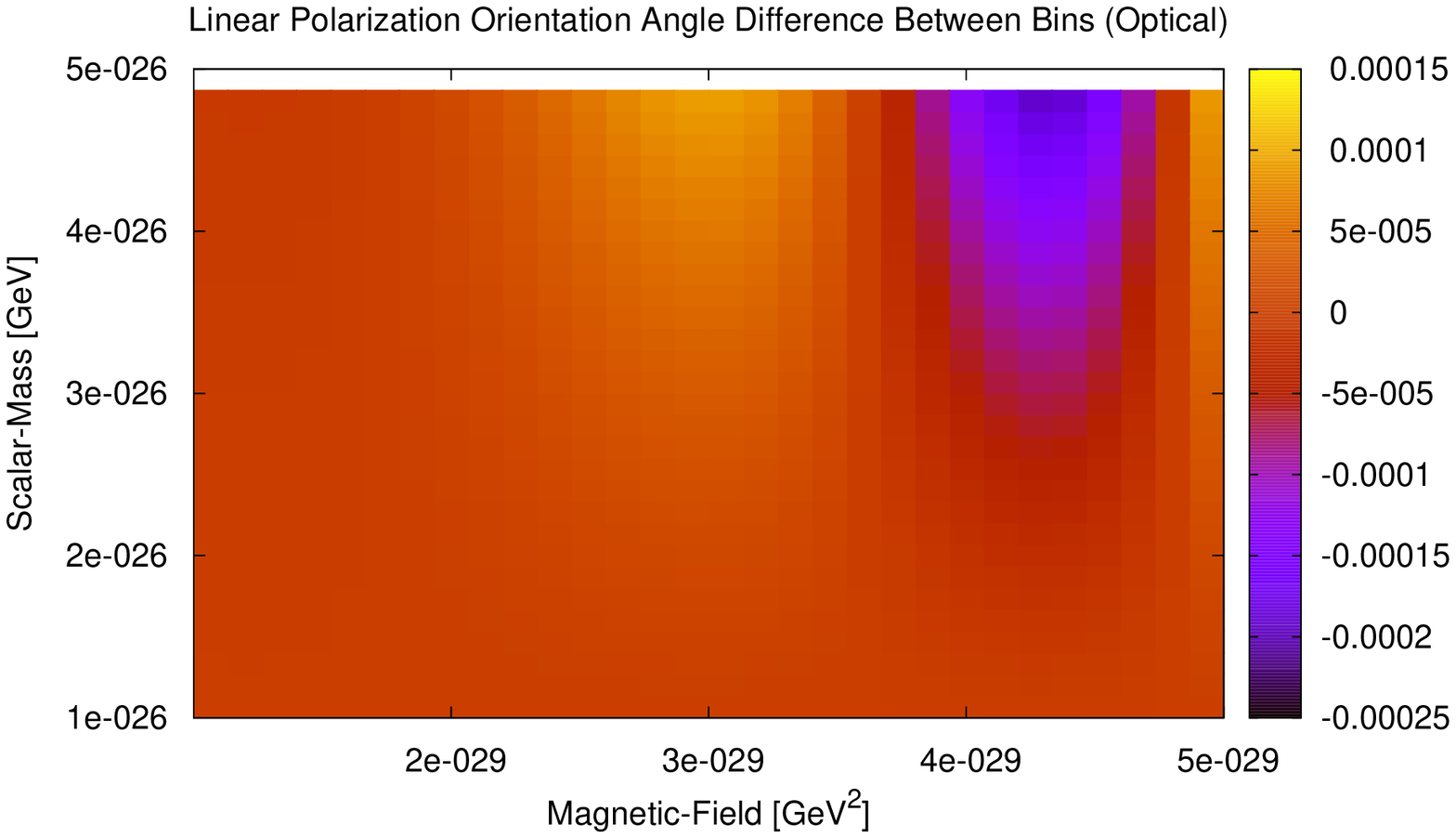}
  & 
     \includegraphics[width=0.5\textwidth]{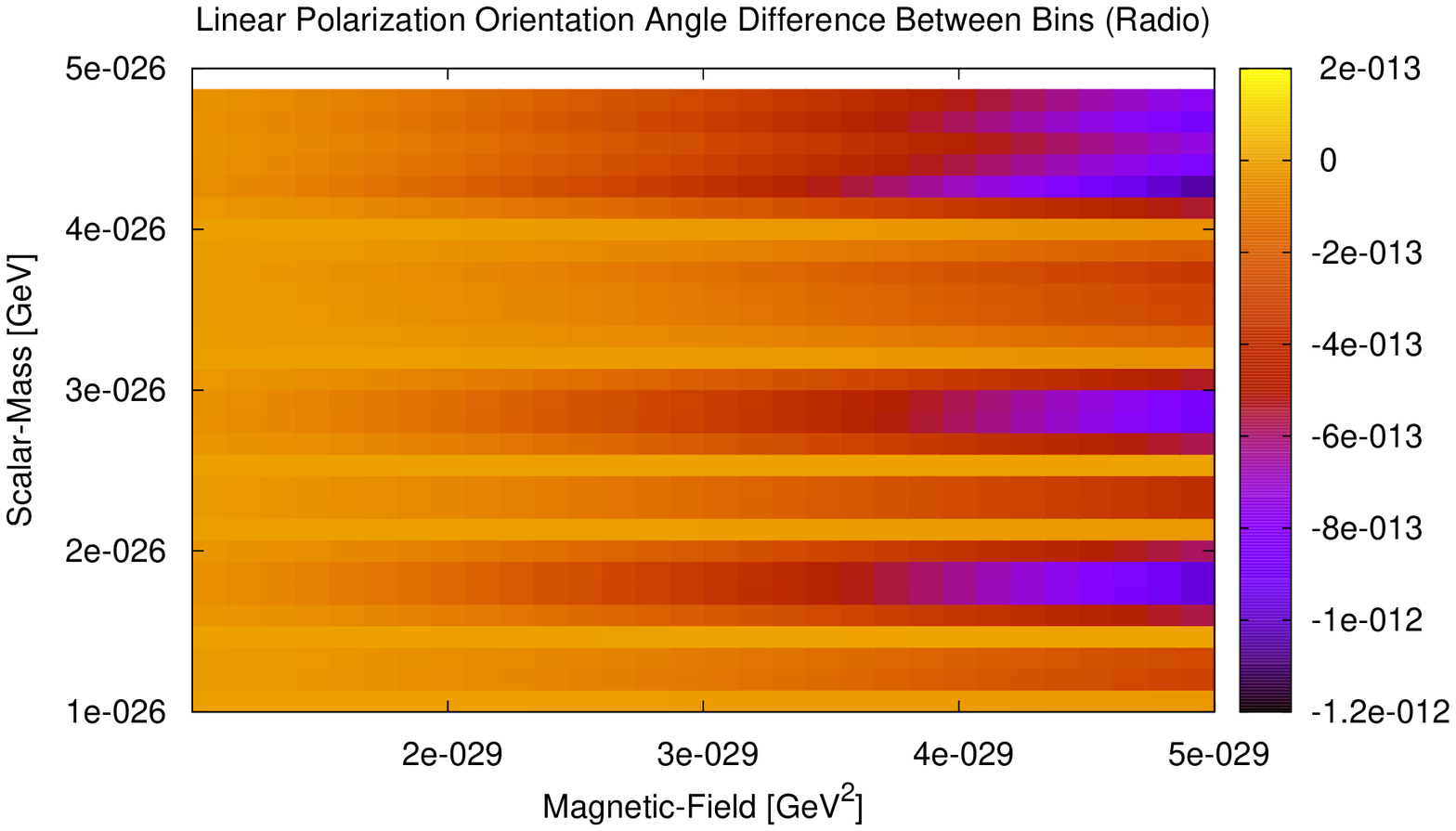}
  \end{tabular}
  \end{center}
  \caption{The  contour plot of the polarization angle change  for case 1) after the light travels over the distance $1$ Gpc in a constant magnetic field given in the horizontal axis  in the optical (left) and radio (right) waves. Here we used the same parameters as in fig. \ref{fig:pol}}.
  \label{fig:dic-1}
\end{figure}
\begin{figure}
  \begin{center}
  \begin{tabular}{cc}
   \includegraphics[width=0.5\textwidth]{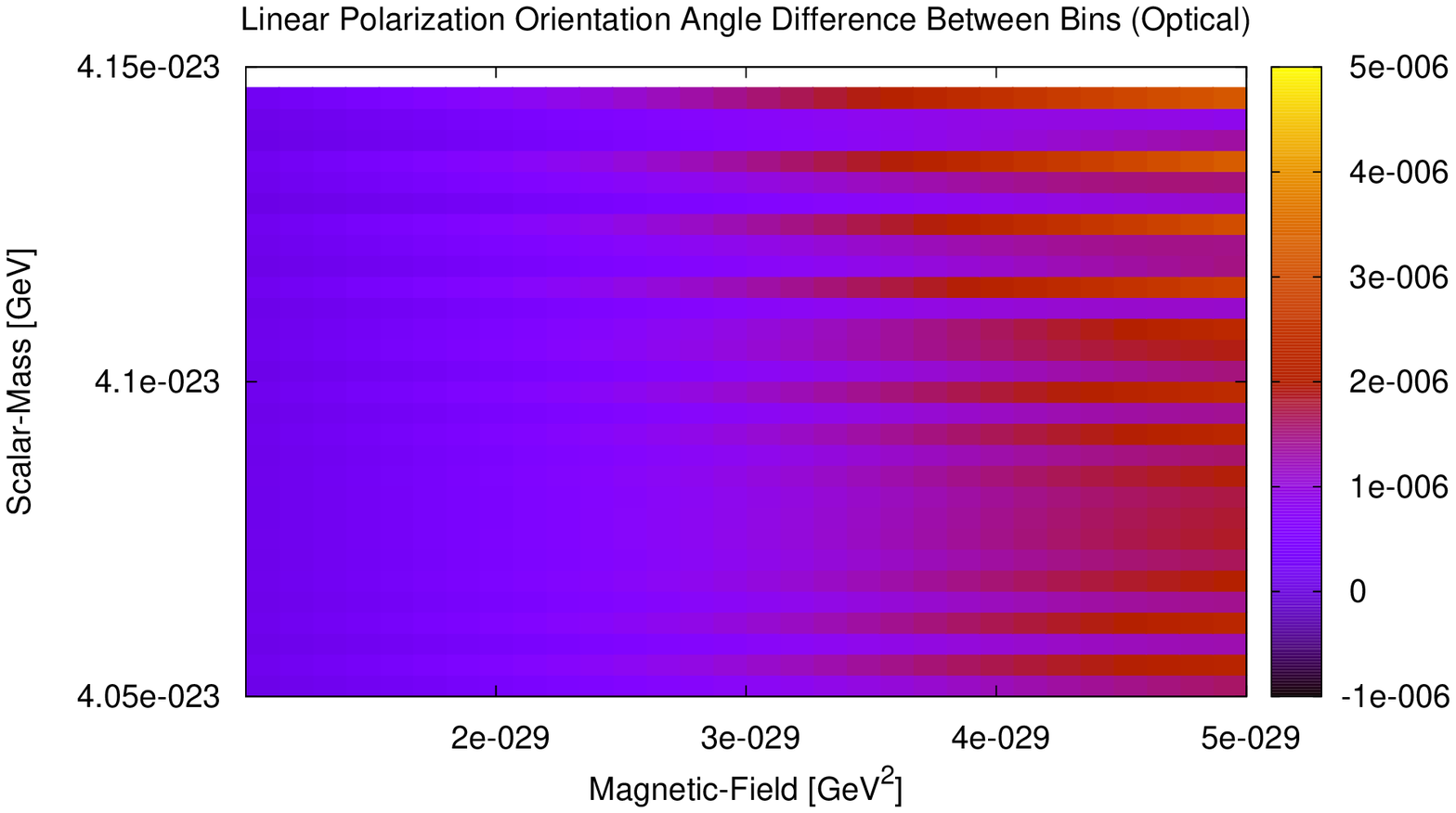}
  & 
     \includegraphics[width=0.5\textwidth]{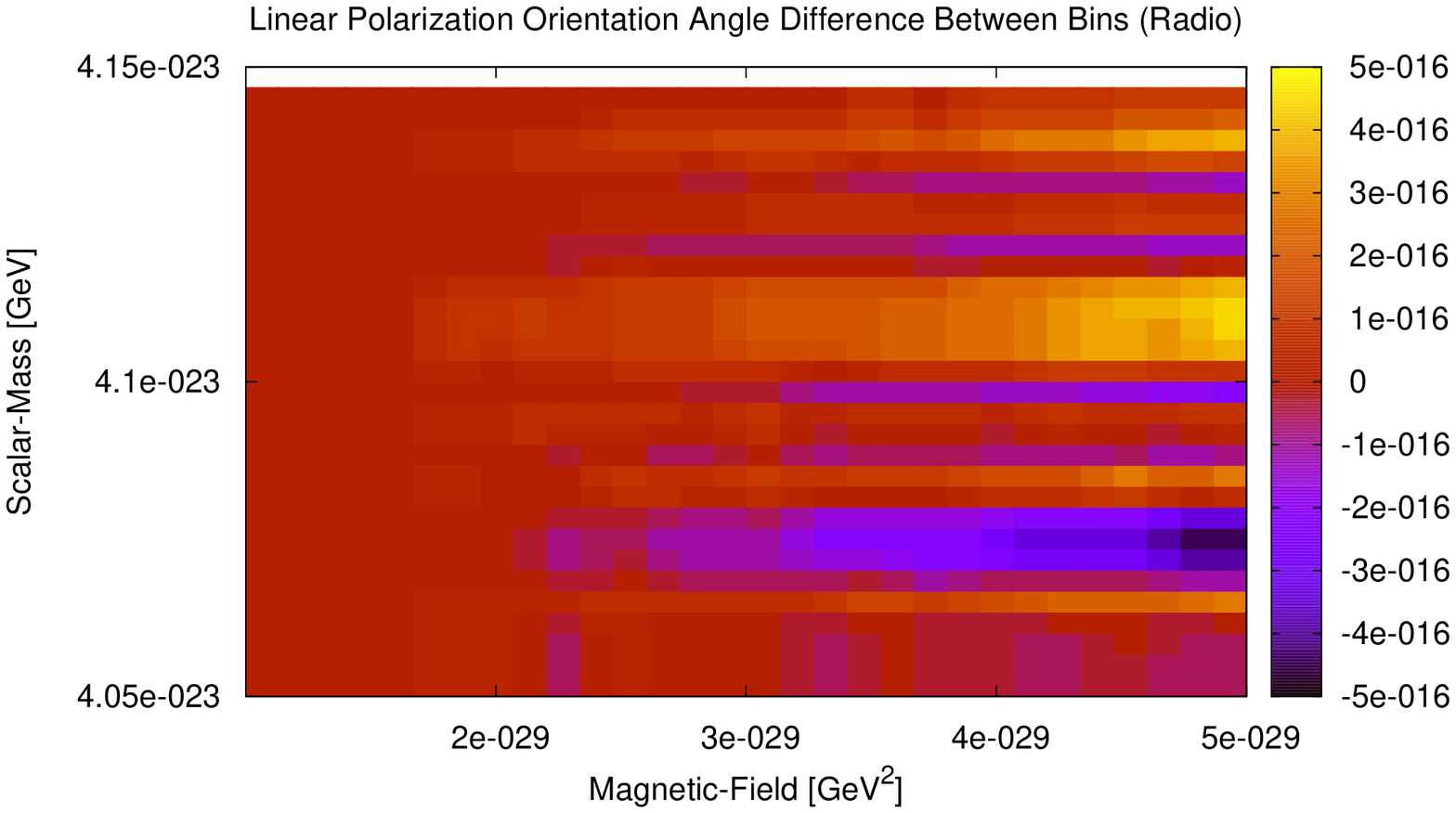}
  \end{tabular}
  \end{center}
  \caption{The  contour plot of the polarization angle change for case 2)  after the light travels over the distance $1$ Gpc in a constant magnetic field given in the horizontal axis.  in the optical (left) and radio (right) waves. Here we used the same parameters as in figure~\ref{fig:poldistance-large}.
}
  \label{fig:dic-2}
\end{figure}

For case 1) \& 2), shown in the previous section \ref{sec:result}, the ALPS can have large mixing and the polarization orientation may change with distance.
The rotation of the polarization angle  is related with the ratio between the relative intensity of the two independent degrees of polarization of light.
The polarization angle at a distance $z$ from the source can be obtained as
\dis{
\psi(z) = \arctan\left[ \frac{|A_\perp(z)|}{|A_\parallel (z)  |} \right]. \label{psiz}
}
 The observation in~\eq{cond_angle} requires that the change of the angle from the source
\dis{
\Delta\psi(z)\equiv \psi(z) - \psi(z=0),
}
satisfies around $\pi/6$ for $ 1$ Gpc distance.

Using Eqs.~(\ref{fields}) and (\ref{psiz}), we find that  the rotations of the polarization angle is given  by
\dis{
\Delta\psi(z) = \arctan\left[ \frac{|A_\perp(0)| \sqrt{ (\cos^2\theta' +\sin^2\theta'\cos\alpha z)^2+\sin^4\theta'\sin^2\alpha z }}{|A_\parallel(0)|\sqrt{ (\cos^2\theta +\sin^2\theta \cos\beta z)^2+\sin^4\theta\sin^2\beta z }  } \right] - \arctan\left[\left|  \frac{A_\perp(0)}{A_\parallel (0)  } \right| \right],\label{angle}
}
where 
\dis{
\alpha \equiv \frac{\mu_+^{'2} - \mu_-^{'2}}{2\omega}, \qquad \rm{and}\qquad \beta \equiv \frac{\mu_+^{2} - \mu_-^{2}}{2\omega}.
}
As you can see in the \eq{angle}, we need unsuppressed mixing angle to have observable effect in the change of the polarization angle.

In figure \ref{fig:dic-1} \& \ref{fig:dic-2}, we show the change of the polarization angle for the case 1) \& 2) shown in the previous section \ref{sec:result}.
We use the same parameters given in the figure \ref{fig:poldistance-large}.   
However we find that the large mixing necessary for the rotation of the polarization angle makes  a linear polarization too large which is inconsistent with observation. At the maximimal case of mixing angle we can obtain the rotation about $\pi/6$ per Gpc.

First, we note that, this shift in orientation with distance is shown by only the restricted data set (aligned to a particular direction), binned over large (z=0.5) distances, with number of entries in each bin being small. Second, the quasar distribution is not uniform (they tend to club together at certain redshifts from earth (both in SGP \& NGP direction). Lastly, there may not be cosmological magnetic fields spanning over several Gpc. Even if there are any such, the directions \& values are supposed to vary with distance. So in this simple model the current (sparse) data on the orientation angle of linear polarization degree of quasars (flocked together in groups which, in turn, are discretely distributed over large distances) may be explained, even if it does so for within slightly lesser distance. Since within that small distance both the restricted dataset and the full one is highly inhomogenous. There could be self-similar variation in the above angle of orientation, that are, presently observed, only in different redshift bins. More data, like LSST \footnote{http://www.lsst.org/lsst/science/development} shall tender us with a more detailed structure of this orientation pattern with minute distances - which may either support our explanation or rule it out. The same stands for this preliminary orientation structure of linear polarization observed at different length scales~\cite{Hutsemekers:2005iz}.

\section{Discussion}
\label{sec:discussion}
We have shown, that for at least ultralight particles of competing nature with slightly different mass, and equal coupling to the  photons may successfully modify the behaviors between the single  ALP and  photon mixing.
The linear polarization observed in the visible and radio waves can have intrinsic origin, however the two ALPs can cancel the possible circular polarization. 

Therefore for two ALPs, we could explain
\begin{enumerate}
\item The linear polarization observed in optical and radio waves from Quasras
\item The absence of circular polarization with observed degree of linear polarization
\end{enumerate} 

Even with two ALPs, however,\\

3. The regular alternance of orientation of polarization in different bins\\

\noindent cannot be produced. Because the large mixing required for the change of rotation angle generates large linear polarization degree and become inconsistent with observation. 

However we note that the redshift dependence of the quasar polarization angles are seen in the selected dataset aligned to a particular direction and binned over large distance. Howeve in the small distance scale 
both the restricted dataset and the full one is highly inhomogenous. For the detailed structure of this orientation pattern with minute distances, which may either support our explanation or rule it out, we need more data such as in the  LSST \footnote{http://www.lsst.org/lsst/science/development}.  The same stands for this preliminary orientation structure of linear polarization observed at different length scales~\cite{Hutsemekers:2005iz}.

We conclude that the single ALP may have problems with the current observation of the polarizations from quasars, 
however two ALPs can overcome these difficulties and can be compatible with the new constraints.

\section*{Acknowledgments}
The authors were supported by Basic Science Research Program through the National Research Foundation of Korea (NRF) funded by the Ministry of Education, Science and Technology (No. 2011-0011083).
The authors acknowledge the Max Planck Society (MPG), the Korea Ministry of
Education, Science and Technology (MEST), Gyeongsangbuk-Do and Pohang
City for the support of the Independent Junior Research Group at the Asia Pacific
Center for Theoretical Physics (APCTP).


\def\prp#1#2#3{Phys.\ Rep.\ {\bf #1} (#3) #2}
\def\rmp#1#2#3{Rev. Mod. Phys.\ {\bf #1} (#3) #2}
\def\anrnp#1#2#3{Annu. Rev. Nucl. Part. Sci.\ {\bf #1} (#3) #2}
\def\npb#1#2#3{Nucl.\ Phys.\ {\bf B#1} (#3) #2}
\def\plb#1#2#3{Phys.\ Lett.\ {\bf B#1} (#3) #2}
\def\prd#1#2#3{Phys.\ Rev.\ {\bf D#1}, #2 (#3)}
\def\prl#1#2#3{Phys.\ Rev.\ Lett.\ {\bf #1} (#3) #2}
\def\jhep#1#2#3{J. High Energy Phys.\ {\bf #1} (#3) #2}
\def\jcap#1#2#3{J. Cosm. and Astropart. Phys.\ {\bf #1} (#3) #2}
\def\zp#1#2#3{Z.\ Phys.\ {\bf #1} (#3) #2}
\def\epjc#1#2#3{Euro. Phys. J.\ {\bf #1} (#3) #2}
\def\ijmp#1#2#3{Int.\ J.\ Mod.\ Phys.\ {\bf #1} (#3) #2}
\def\mpl#1#2#3{Mod.\ Phys.\ Lett.\ {\bf #1} (#3) #2}
\def\apj#1#2#3{Astrophys.\ J.\ {\bf #1} (#3) #2}
\def\nat#1#2#3{Nature\ {\bf #1} (#3) #2}
\def\sjnp#1#2#3{Sov.\ J.\ Nucl.\ Phys.\ {\bf #1} (#3) #2}
\def\apj#1#2#3{Astrophys.\ J.\ {\bf #1} (#3) #2}
\def\ijmp#1#2#3{Int.\ J.\ Mod.\ Phys.\ {\bf #1} (#3) #2}
\def\apph#1#2#3{Astropart.\ Phys.\ {\bf B#1}, #2 (#3)}
\def\mnras#1#2#3{Mon.\ Not.\ R.\ Astron.\ Soc.\ {\bf #1} (#3) #2}
\def\nat#1#2#3{Nature (London)\ {\bf #1} (#3) #2}
\def\apjs#1#2#3{Astrophys.\ J.\ Supp.\ {\bf #1} (#3) #2}
\def\aipcp#1#2#3{AIP Conf.\ Proc.\ {\bf #1} (#3) #2}


\end{document}